\documentclass[12pt]{article}
\usepackage{epsfig,graphicx}
\usepackage{ch2005}
\usepackage{amssymb}
\usepackage{amsmath,amsfonts}

\def\beq{\begin{equation}}
\def\eeq#1{\label{#1}\end{equation}}
\def\eeqn{\end{equation}}

\def\beqa{\begin{eqnarray}}
\def\eeqa#1{\label{#1}\end{eqnarray}}
\def\eeqan{\end{eqnarray}}

\def\Dslash{\not{\hbox{\kern-4pt $D$}}}
\def\dslash{\not{\hbox{\kern-2pt $\del$}}}

\newcommand{\tev}{\ensuremath{\mathrm{\,Te\kern -0.1em V}}\xspace}
\newcommand{\gev}{\ensuremath{\mathrm{\,Ge\kern -0.1em V}}\xspace}
\newcommand{\mev}{\ensuremath{\mathrm{\,Me\kern -0.1em V}}\xspace}
\newcommand{\kev}{\ensuremath{\mathrm{\,ke\kern -0.1em V}}\xspace}
\newcommand{\ev}{\ensuremath{\mathrm{\,e\kern -0.1em V}}\xspace}
\newcommand{\gevc}{\ensuremath{{\mathrm{\,Ge\kern -0.1em V\!/}c}}\xspace}
\newcommand{\mevc}{\ensuremath{{\mathrm{\,Me\kern -0.1em V\!/}c}}\xspace}
\newcommand{\gevcc}{\ensuremath{{\mathrm{\,Ge\kern -0.1em V\!/}c^2}}\xspace}
\newcommand{\mevcc}{\ensuremath{{\mathrm{\,Me\kern -0.1em V\!/}c^2}}\xspace}

\def\mus  {\ensuremath{\rm \,\mus}\xspace}

\def\mus        {\ensuremath{\,\mu{\rm s}}\xspace}    

\begin{document}


\Title{FADC Pulse Reconstruction Using a Digital Filter for the MAGIC Telescope}
\bigskip


%
\label{BartkoStart}

%
\author{ H. Bartko\,$^{a}$\index{Bartko, H.}, M. Gaug\,$^{b}$\index{Gaug, M.}, A. Moralejo\,$^{c}$\index{Moralejo, A.}, N. Sidro\,$^{b}$\index{Sidro, N.}  for the MAGIC collaboration }

%
\address{(a) Max-Planck-Institute for Physics, Munich, Germany \\ 
(b) Institut de Fisica d Altes Energies, Bellaterra, Spain \\
(c) University and INFN Padova, Italy
}


\makeauthor\abstracts{
Presently, the MAGIC telescope uses a 300~MHz FADC system to sample the
transmitted and shaped signals from the captured Cherenkov light of air
showers.
We describe a method of Digital Filtering of the FADC samples to extract the charge and the 
arrival time of the signal: Since the pulse shape is dominated by 
the electronic pulse shaper, a numerical fit can be applied to the FADC samples taking the noise 
autocorrelation into account. The achievable performance of the digital filter is 
presented and compared to other signal reconstruction algorithms.
}




\section{Introduction} 

The purpose of the MAGIC Telescope \cite{MAGIC-commissioning} is the observation of
 high energy gamma radiation from celestial objects. When the gamma quanta hit
 the earth atmosphere they initiate a cascade of photons, electrons and
 positrons. The latter radiate short flashes of Cherenkov light which can be recorded
 by a Cherenkov telescope. The FWHM of the pulses is about 2\,ns.

\begin{figure}[h!]
\begin{center}
\includegraphics[totalheight=6cm]{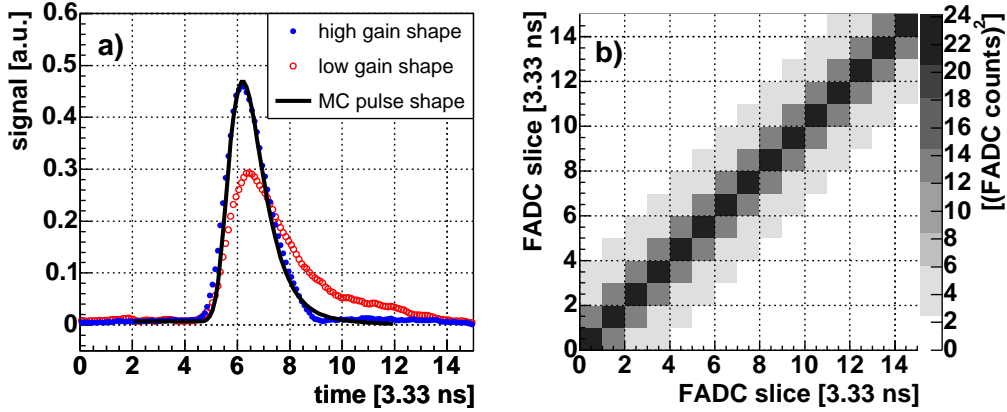}
\end{center}
\caption[Reconstructed pulse shapes]{\small \it a) Average normalized reconstructed high and low gain pulse shapes and pulse shape implemented in the MC simulations \cite{MC-Camera}.
b) Noise autocorrelation matrix $\boldsymbol{B}$ for open camera and averaged over all pixels (galactic telescope pointing).} 
\label{fig:pulse_shapes}
\end{figure}

In order to sample this pulse shape with the 300 MSamples/s FADC system \cite{Magic-DAQ}, the original pulse is folded with a 
stretching function leading to a FWHM  greater than 6\,ns. To increase the dynamic range of the MAGIC FADCs the signals are split into two branches with gains differing by a factor 10. Figure~\ref{fig:pulse_shapes}a) shows a typical average of identical signals.



In order to discriminate the small signals from showers in the energy range below 100 GeV against the light of the night sky (LONS) the highest possible signal to noise ratio, 
signal reconstruction resolution and a small bias are important.

Monte Carlo (MC) based simulations predict different time structures for gamma and hadron induced shower 
images as well as for images of single muons \cite{muon_rejection}. An accurate arrival time determination may therefore improve 
the separation power of gamma events from the background events. Moreover, the timing information may be 
used in the image cleaning to discriminate between pixels whose signal belongs to the shower and pixels 
which are dominated by randomly timed background noise.

\section{Digital Filter}

\begin{figure}[htp]
\begin{center}
\includegraphics[totalheight=6cm]{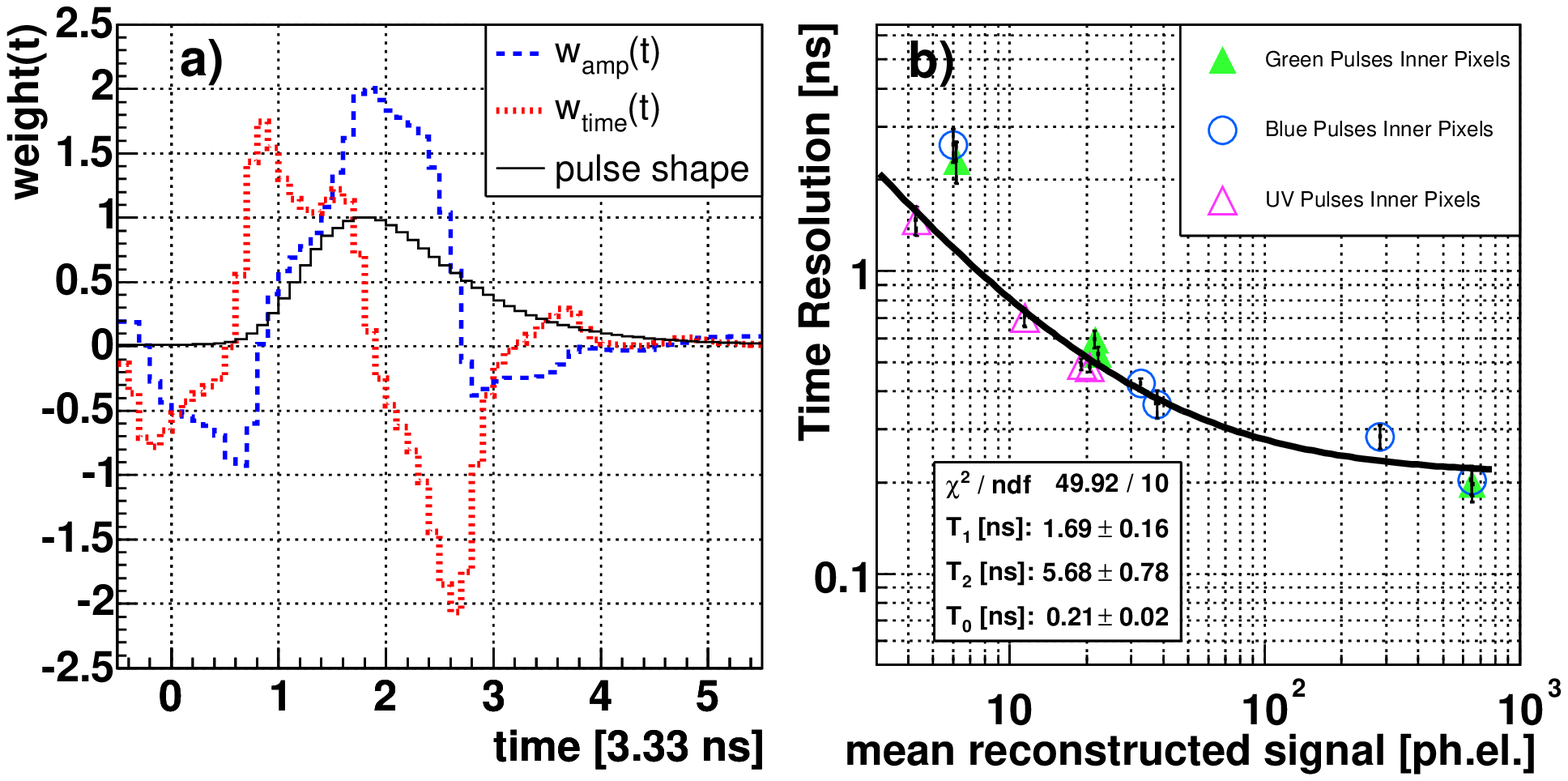}
\end{center}
\caption[Noise autocorrelation average all pixels.]{\small \it a) Amplitude weights $w_{\mathrm{amp}}(t_0) \ldots w_{\mathrm{amp}}(t_5)$ for a window size of 6 FADC slices for the pulse shape used in the MC simulations, see text. b) Timing resolution for calibration LED pulses \cite{MAGIC-calibration}.} 
\label{fig:noise_autocorr_allpixels}
\end{figure}



The goal of the digital filtering method \cite{OF94,OF77} is to optimally reconstruct from FADC samples the amplitude and arrival time of a signal whose shape is known. Thereby, the noise contributions to the amplitude and arrival time reconstruction are minimized.

For the digital filtering method, three assumptions have to be made:

\begin{itemize}
\item{The normalized signal shape has to be always constant.}
\item{The noise properties must be constant, especially independent of the signal amplitude.}
\item{The normalized noise auto-correlation has to be constant.}
\end{itemize}

Due to the artificial pulse stretching by about 6\,ns on the receiver board all three assumptions are fullfilled to a good approximation. For a more detailed discussion see \cite{OF_Magic_NIM}.

Let $g(t)$ be the normalized signal shape, $E$ the signal amplitude and $\tau$ the time shift 
between the physical signal and the predicted signal shape. Then the time dependence of the signal, $y(t)$, is given by $y(t)=E \cdot  g(t-\tau) + b(t) \ ,$ where $b(t)$ is the time-dependent noise contribution. For small time shifts $\tau$ (usually smaller than 
one FADC slice width), 
the time dependence can be linearized. Discrete
measurements $y_i$ of the signal at times $t_i \ (i=1,...,n)$ have the form $y_i=E \cdot g_i- E\tau \cdot \dot{g}_i + O(\tau^2) +b_i$, where $\dot{g}(t)$ is the time derivative of the signal shape.



The correlation of the noise contributions at times $t_i$ and $t_j$ can be expressed in the noise autocorrelation matrix $\boldsymbol{B}$: $B_{ij} = \langle b_i b_j \rangle - \langle b_i \rangle \langle b_j\rangle$.  Figure \ref{fig:noise_autocorr_allpixels} shows the noise 
autocorrelation matrix for an open camera. It is dominated by LONS pulses shaped to 6\,ns.



The signal amplitude $E$, and the product $E \tau$ of amplitude and time shift, can be estimated from the given FADC measurements $\boldsymbol{y} = (y_1, ... ,y_n)$ by minimizing the deviation of the measured FADC slice contents from the 
known pulse shape with respect to the known noise auto-correlation: $\chi^2(E, E\tau) = (\boldsymbol{y} - E
\boldsymbol{g} - E\tau \dot{\boldsymbol{g}})^T \boldsymbol{B}^{-1} (\boldsymbol{y} - E \boldsymbol{g}- E\tau \dot{\boldsymbol{g}}) + O(\tau^2)$ (in matrix form). This leads to the following solution for  $\overline{E}$ and $\overline{E\tau}$:


\begin{equation}
\overline{E} = \boldsymbol{w}_{\text{amp}}^T(t_{\text{rel}}) \boldsymbol{y} + O(\tau^2) \quad \mathrm{with} \quad 
        \boldsymbol{w}_{\text{amp}}(t_{\text{rel}}) 
        = \frac{ (\dot{\boldsymbol{g}}^T\boldsymbol{B}^{-1}\dot{\boldsymbol{g}}) \boldsymbol{B}^{-1} \boldsymbol{g} -(\boldsymbol{g}^T\boldsymbol{B}^{-1}\dot{\boldsymbol{g}})  \boldsymbol{B}^{-1} \dot{\boldsymbol{g}}}  
        {(\boldsymbol{g}^T \boldsymbol{B}^{-1} \boldsymbol{g})(\dot{\boldsymbol{g}}^T\boldsymbol{B}^{-1}\dot{\boldsymbol{g}}) -(\dot{\boldsymbol{g}}^T\boldsymbol{B}^{-1}\boldsymbol{g})^2 } \ ,
\end{equation}

\begin{equation}
\overline{E\tau} = \boldsymbol{w}_{\text{time}}^T(t_{\text{rel}}) \boldsymbol{y} + O(\tau^2) \quad 
        \mathrm{with} \quad \boldsymbol{w}_{\text{time}}(t_{\text{rel}})
        = \frac{ ({\boldsymbol{g}}^T\boldsymbol{B}^{-1}{\boldsymbol{g}}) \boldsymbol{B}^{-1} \dot{\boldsymbol{g}} -(\boldsymbol{g}^T\boldsymbol{B}^{-1}\dot{\boldsymbol{g}})  \boldsymbol{B}^{-1} {\boldsymbol{g}}}  
        {(\boldsymbol{g}^T \boldsymbol{B}^{-1} \boldsymbol{g})(\dot{\boldsymbol{g}}^T\boldsymbol{B}^{-1}\dot{\boldsymbol{g}}) -(\dot{\boldsymbol{g}}^T\boldsymbol{B}^{-1}\boldsymbol{g})^2 } \ ,
\end{equation}

where $t_{\text{rel}}$ is the relative phase between $g(t)$ and the FADC clock. Thus $\overline{E}$ and $\overline{E\tau}$ are given by a weighted sum of the discrete measurements $y_i$ with the weights for the amplitude, $w_{\text{amp}}(t_{\text{rel}})$, and time shift, $w_{\text{time}}(t_{\text{rel}})$, plus $O(\tau^2)$. To reduce $O(\tau^2)$ the fit can be iterated using $g(t_1=t-\tau)$ and the weights $w_{\mathrm{amp/time}}(t_{\text{rel}}+\tau)$ \cite{OF94,OF_Magic_NIM}. Figure \ref{fig:noise_autocorr_allpixels} a) shows the amplitude and timing weights for the MC pulse shape. 
The first weight $w_{\mathrm{amp/time}}(t_0)$ is plotted as a function of $t_{\text{rel}}$ in the range $[-0.5,0.5[ \ T_{\text{ADC}}$, the second weight in the range $[0.5,1.5[ \ T_{\text{ADC}}$ and so on. 







The expected contributions of the noise to the error of the estimated amplitude and timing only depend on the the shape $g(t)$, and the noise auto-correlation $B$. Analytic expressions can be found in references \cite{OF94,OF_Magic_NIM}.

\section{Performance and Discussion}

Figure \ref{fig:noise_autocorr_allpixels}b) shows the measured timing resolution for different calibration LED pulses as a function of the mean reconstructed pulse charge. For signals of 10 photo-electrons the timing resolution is as good as 700\,ps, for very large signals a timing resolution of about 200\,ps can be achieved.

\begin{figure}[htp]
\begin{center}
  \includegraphics[totalheight=6cm]{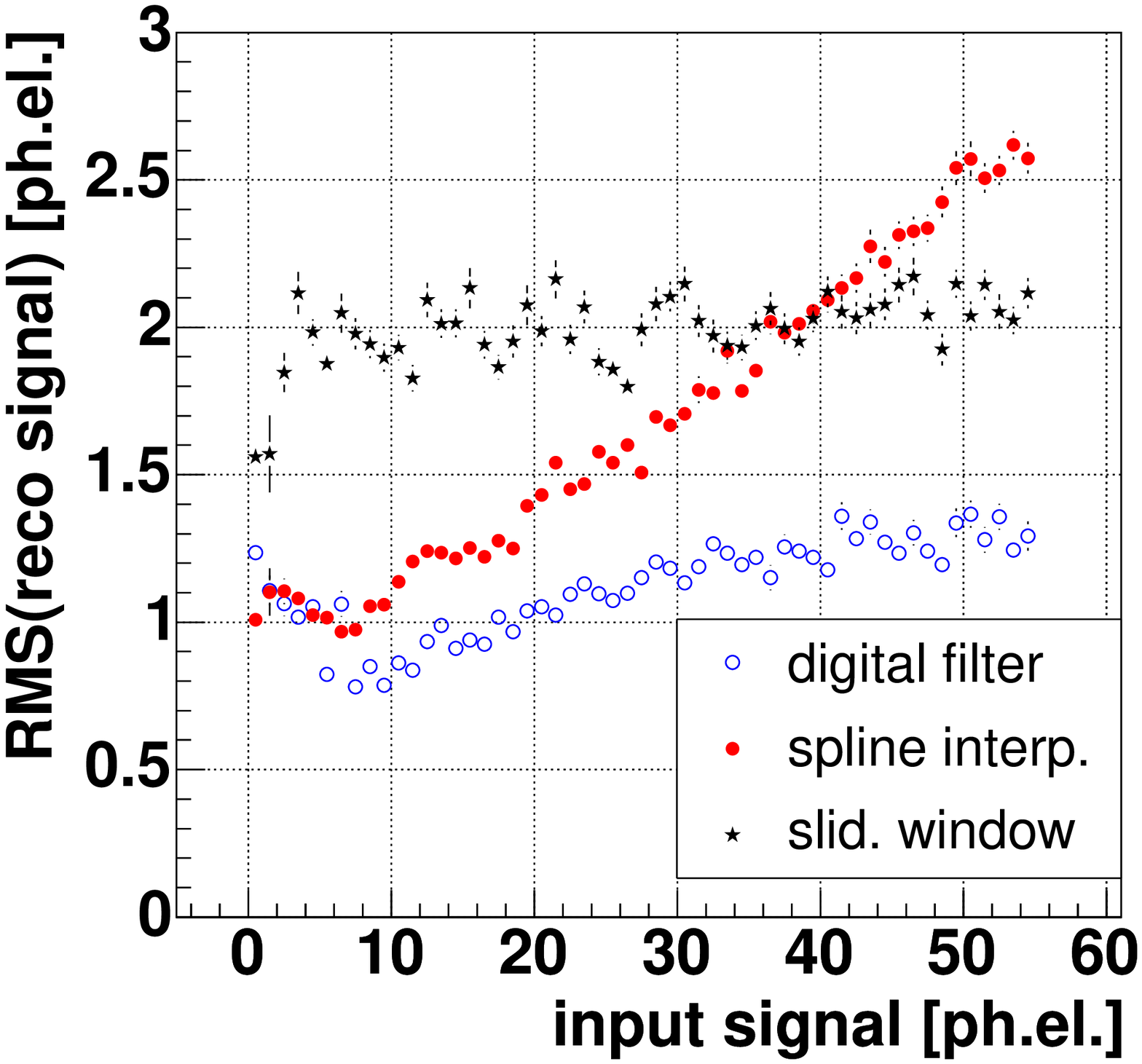}
  \includegraphics[totalheight=6cm]{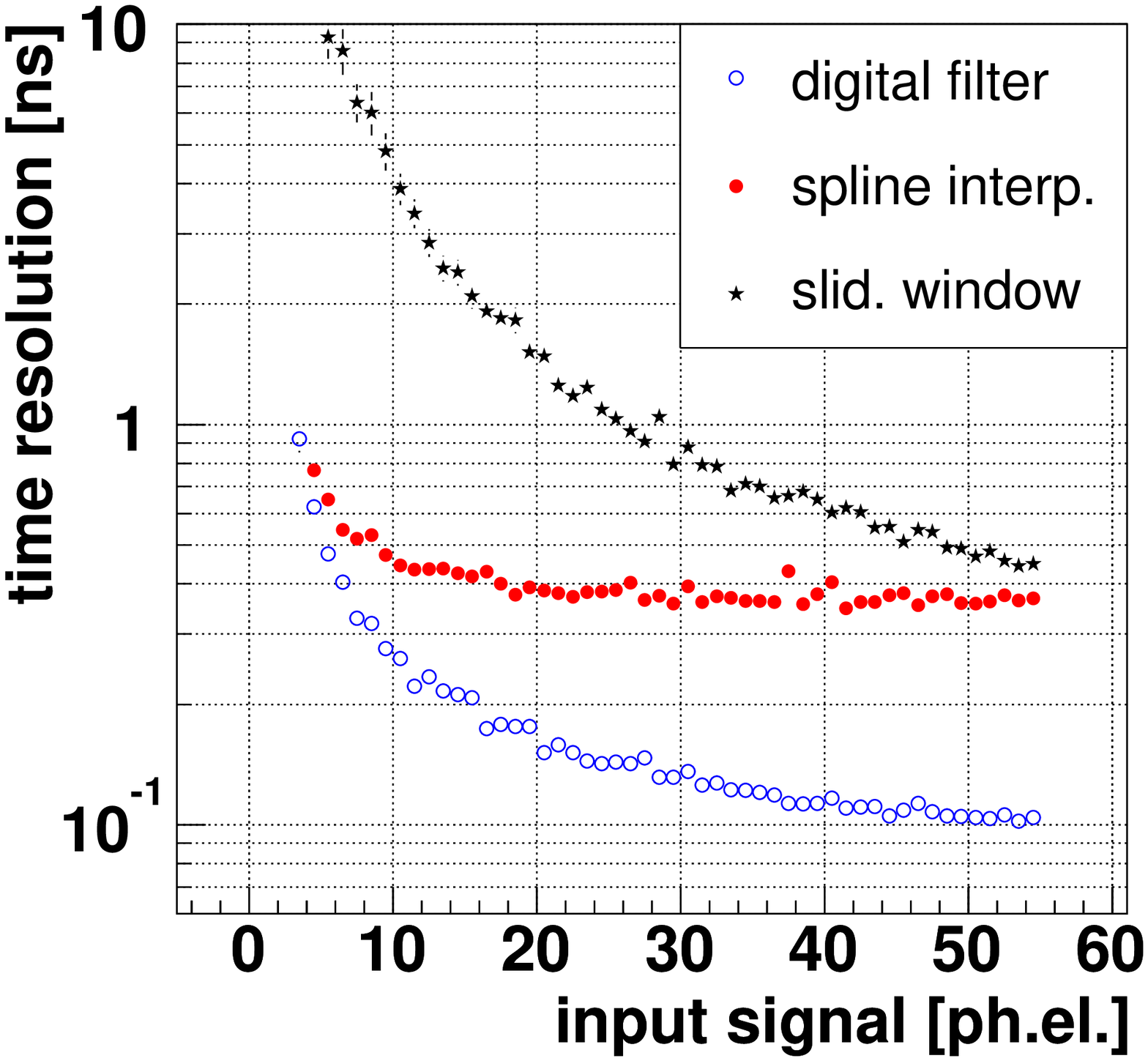}
\caption[Amplitude / timing resolution for 10 LED UV]{%
\small \it Charge and arrival time resolution as a function of the input pulse height for MC simulations for the Digital Filter, a cubic spline interpolation (charge= spline integral, time=half maximum position) and a sliding window of 6 FADC slices (charge=samples sum, time = pulse barycenter) \cite{OF_Magic_NIM}.} 
\label{fig:dfsketch}
\end{center}
\end{figure}

Figure \ref{fig:dfsketch} shows the charge and arrival time resolution as a function of the input pulse height for MC simulations (no PMT time spread and no gain fluctuations) assuming an extra-galactic background for different signal extraction algorithms. The digital filter yields the best charge and timing resolution of the studied algorithms  \cite{OF_Magic_NIM}.

For known constant signal shapes and noise auto-correlations
 the digital filter yields the best theoretically achievable signal and timing
 resolution. Due to the pulse shaping of the Cherenkov signals the algorithm can be applied to reconstruct their charge and
 arrival time\pagebreak, although there are some fluctuations of the pulse shape and
 noise behavior. The digital filter reduces the noise contribution to the error of the reconstructed signal. Thus it is possible to
 lower the image cleaning levels and the analysis energy threshold \cite{OF_Magic_NIM}. The timing
 resolutions is as good as a few hundred ps for large signals.



\appendix

\section*{Acknowledgements}

The authors thank F. Goebel, Th. Schweizer and W. Wittek for discussions and suggestions. 






\begin{thebibliography}{1}

\bibitem{MAGIC-commissioning}
C.~Baixeras et~al. (MAGIC Collab.),
\newblock Nucl. Instrum. Meth. {\bf A518} (2004) 188.

\bibitem{Magic-DAQ}
F.~Goebel et~al. (MAGIC Collab.),
\newblock in {\em Proceedings of the 28th ICRC}, Tokyo, 2003. 

\bibitem{MC-Camera}
A.~Moralejo et~al.,
\newblock {\em MC Simulations for the MAGIC Telescope},
\newblock In preparation.

\bibitem{OF94}
W.~E. Cleland and E.~G. Stern,
\newblock Nucl. Instrum. Meth. {\bf A338} (1994) 467.

\bibitem{OF77}
A.~Papoulis,
\newblock {\em Signal analysis},
\newblock McGraw-Hill, 1977.

\bibitem{MAGIC-calibration}
T.~Schweizer et~al.,
\newblock IEEE Trans. Nucl. Sci. {\bf 49} (2002) 2497.

\bibitem{muon_rejection}
R.~Mirzoyan et~al., 
\newblock to be published in proceedings of the conference {\em Towards a Network of Atmospheric Cherenkov Detectors VII}, 27-29 April 2005 Palaiseau, France. 

\bibitem{OF_Magic_NIM}
H.~Bartko et~al.,
\newblock In preparation, to be submitted to Nucl. Inst. Meth.


\end{thebibliography}
\bibliographystyle{bibstyle} 

\end{document}